\documentstyle[11pt,aaspp4,flushrt]{article}

\begin{document}

\title{Radio Pulses along the Galactic Plane}
\author{David J. Nice}
\affil{Joseph Henry Laboratories and Physics Department \\
Princeton University, Princeton, NJ 08544}

\bigskip
\bigskip

\centerline{Submitted to the {\it Astrophysical Journal}}
\centerline{Revised 28 September 1998}
\centerline{Accepted 5 October 1998}

\bigskip
\bigskip

\begin{abstract}

We have surveyed 68\,deg$^2$ along the Galactic Plane for single, dispersed
radio pulses.  Each of 3027 independent pointings was observed for 68\,s using
the Arecibo telescope at 430\,MHz.  Spectra were collected at intervals of
0.5\,ms and examined for pulses with duration 0.5 to 8\,ms.  Such single pulse
analysis is the most sensitive method of detecting highly scattered or highly
dispersed signals from pulsars with large pulse-to-pulse intensity variations.
A total of 36 individual pulses from five previously known pulsars were
detected, along with a single pulse not associated with a previously known
source.  Follow-up observations discovered a pulsar, PSR\,J1918+08, from which
the pulse originated.  This pulsar has period 2.130\,s and dispersion measure
30\,pc\,cm$^{-3}$, and has been seen to emit single pulses with strength up to
8 times the average.

\end{abstract}

\keywords{Pulsars: General --- Pulsars: Individual (PSR\,J1918+08)}

\section{Introduction}\label{sec:intro}

Pulsars were first discovered by the examination of time series of radio total
power measurements (chart recorder output) for pulses (\cite{hbp+68}).  Since
pulsar signals are expected to be periodic, modern pulsar searches have
eschewed this single-pulse scheme in favor of searches for periodic signals,
using Fourier or periodogram techniques, combined with a suitable
de-dispersion algorithm.  Use of a periodic search scheme typically leads to a
large increase in sensitivity, as it incorporates many pulses from a single
source, and the signal-to-noise ratio scales as $n^{1/2}$ where $n$ is the
number of pulses observed.

Under certain circumstances periodic schemes fail.  Very long period pulsars
may be missed because there are few pulses in an observation, and
pulse-to-pulse intensity variations weaken the periodic signal; in addition,
terrestrial radio frequency interference tends to obscure weak, long period
astrophysical signals.  Very short period pulsars may be missed due to
broadening by finite instrumental time resolution, or by dispersion or
scattering in the interstellar medium.  Under these conditions some
individual pulses may still be detectable.  Pulsars are well known to exhibit
pulse-to-pulse intensity variations, and at least two pulsars exhibit ``giant
pulses,'' individual pulses with flux densities hundreds or thousands times
greater than average.  If the time resolution is limited to a few milliseconds by
instrumentation or by interstellar propagation effects, the periodic signal of
a fast pulsar will not be detectable, yet a sufficiently strong single pulse,
even if broadened to a duration of several milliseconds, will still be
detectable.  Indeed, the first detection of time variable radio signals from
the Crab pulsar consisted of only giant pulses, because the periodic signal
was faster than the instrumental time resolution (\cite{sr68}).

Motivated by the possibility of detecting hitherto unknown pulsars, as well as
expanding the parameter space searched for transient radio signals, we have
reanalyzed a set of high sensitivity pulsar search data collected along the
Galactic plane.  The original analysis of these data was presented in Nice,
Fruchter, \& Taylor (1995, Paper I)\nocite{nft95}.  In that work, conventional
periodic search techniques detected fourteen pulsars in the portion of the
data reanalyzed here.  This portion of the sky has been searched once before
for single pulses, with substantially poorer time resolution (\cite{pt79}).

\section{Observations}

\begin{figure}
\center{\ \plotone{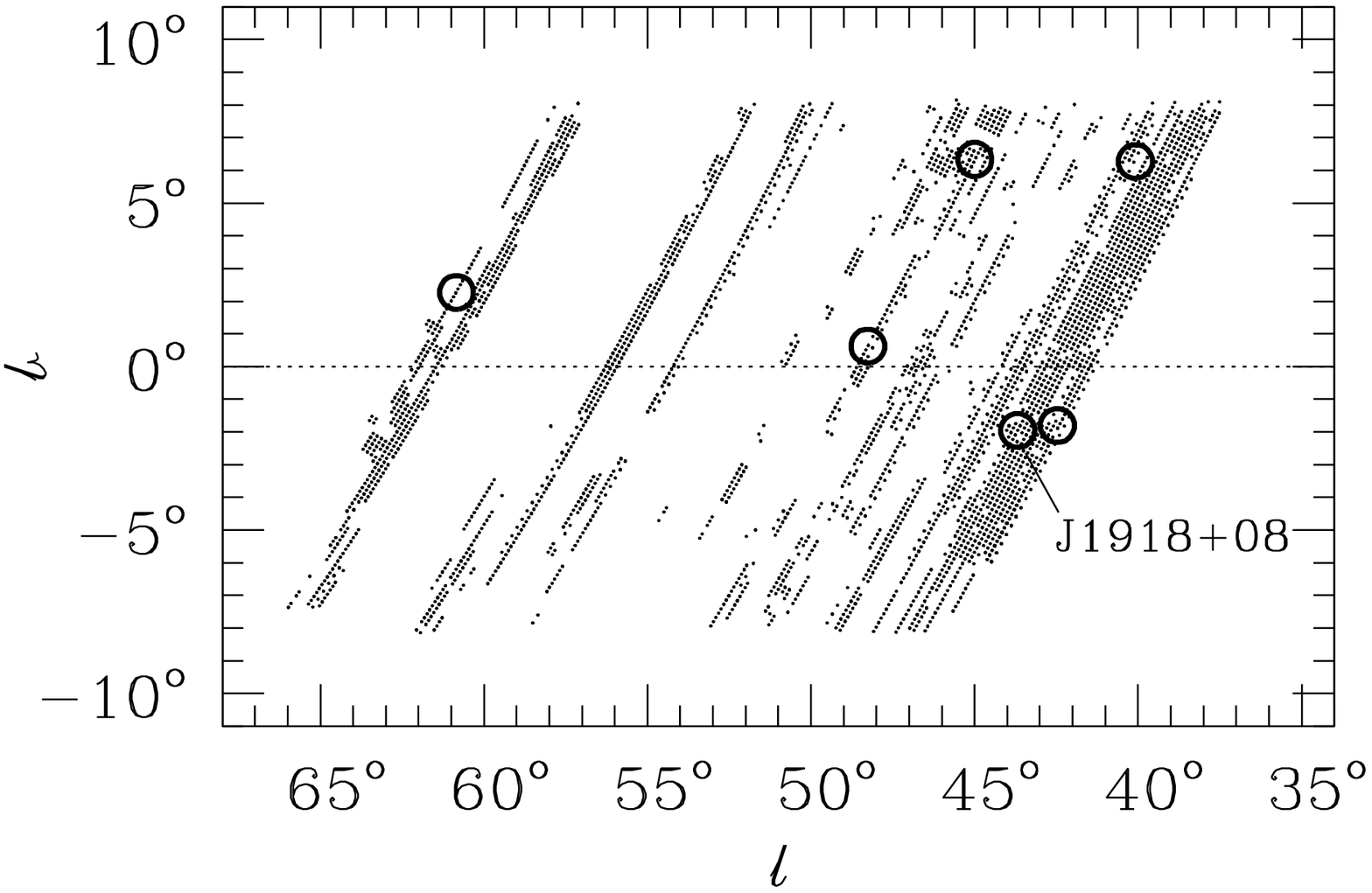}}
\caption{\label{fig:map}
Coverage of the search in Galactic coordinates.  Search pointings are
indicated by small dots.  Detected pulsars are indicated by large open
circles.  The newly discovered pulsar J1918+08 is indicated by name.
}
\end{figure}

Observations were made with the 305~m radio telescope at Arecibo, Puerto Rico,
at 430\,MHz at several epochs in 1990 and 1991.  Details of the observations
are given in Paper\,I.  Briefly, pointings were selected from a $9'\times 9'$
grid at Galactic latitudes $|b|<8^\circ$ and longitudes $36^\circ<l<70^\circ$.
Each pointing was observed for 68~s.  A correlation spectrometer synthesized
128 spectral channels across 10~MHz passbands in right and left circular
polarization.  Spectra were measured at intervals of 516.625\,$\mu$s, for a
total of 131,072 spectra per pointing.  Spectra from opposite polarizations
were summed, reduced to four bits, and written to tape for off-line
processing.  The original survey covered more than 10,000 pointings.  Good
quality data from 3027 pointings were reanalyzed for this work
(figure~\ref{fig:map}).

\section{Analysis}

Each pointing was independently searched for pulses with duration 0.5
to 8\,ms.  To search for 0.5\,ms pulses, time series were
generated at 256 trial dispersion measures by summing spectral channels
after delaying them relative to 
one another to compensate for dispersion.  The trial dispersion measures, from
0 to 127\,pc\,cm$^{-3}$ at intervals of 0.5\,pc\,cm$^{-3}$, were equivalent to
delays between the highest and lowest spectral channels of 0 to 255 sample
intervals.  At dispersion measures above 127\,pc\,cm$^{-3}$, dispersion
smearing was more than 1\,ms, well above the resolution needed to distinguish
0.5\,ms pulses.  To search for pulses with higher dispersion measures, as
well as longer durations, adjacent
sets of 2, 4, 8, and 16 spectra were
summed, and the de-dispersion algorithm repeated on each set.  The 
time resolutions and dispersion measure ranges resulting from this
procedure are listed in
table~\ref{tab:param}.  

Each de-dispersed time series was normalized and searched for points
substantially higher than the mean.  Comparisons of detections at
adjacent times and dispersion measures were made to find the time, pulse
duration, and dispersion measure that gave the highest signal-to-noise ratio
for a candidate pulse.

Candidate pulses with dispersion measures indistinguishable from zero were
rejected, as they arise from terrestrial radio frequency interference.
Typically there were one or two such pulses per pointing.  The lower limits on
dispersion measure listed in table~\ref{tab:param} for each time resolution
were found to be practical minimum values for distinguishing dispersed and
non-dispersed signals.

Despite the rejection of non-dispersed pulses, manual inspection of power
series and spectra around many remaining candidate pulses suggested that radio
frequency interference was the source of a substantial number of spurious
detections.  Practical {\it a posteriori} cutoffs in signal-to-noise ratio
(S/N) were selected for each time resolution (table~\ref{tab:param}) such that
all signals above the quoted S/N thresholds were of astrophysical origin
(\S\ref{sec:detections}).  In each case there were numerous candidate pulses
close to but below the threshold.  For example, the search with 1.1\,ms
resolution was assigned a S/N threshold of 9.  There were two candidates with
$8<{\rm S/N}<9$, and twelve candidates with $7<{\rm S/N}<8$.  Follow-up
observations of a large number of low signal-to-noise candidates was not
deemed practical, since most or all were likely spurious detections
caused by statistical fluctuations or terrestrial interference.

\subsection{Sensitivity}

The system temperature and telescope gain varied substantially from
observation to observation due to the spillover characteristics of the
telescope and to variations in sky temperature along the Galactic plane.
These phenomena are discussed in Paper I.  To simplify analysis of the current
survey, we adopt the median system equivalent noise value from that work,
\begin{equation}
N = \frac{T_{\rm sys}}{G(n_p B t)^{1/2}} = 0.156\,{\rm Jy}\,n^{-1/2},
\end{equation}
where $n$ is the number of summed samples, and $T_{\rm sys}$, $G$, $n_p$, $B$,
and $t$ are the system temperature, gain, number of polarizations, bandwidth,
and integration time, respectively.

We use the term ``pulse strength'' to refer to the flux density integrated
over the duration of the pulse, that is, energy per unit area per unit
bandwidth.  It is convenient to work in units of
$10^{-4}$\,Jy\,s=$10^{-30}$\,J\,m$^{-2}$\,Hz$^{-1}$.  
Table~\ref{tab:param} gives the minimum detectable pulse strength for each
time resolution analyzed in the search.

\subsection{Detections}\label{sec:detections}

\begin{figure}
\center{\ \plotone{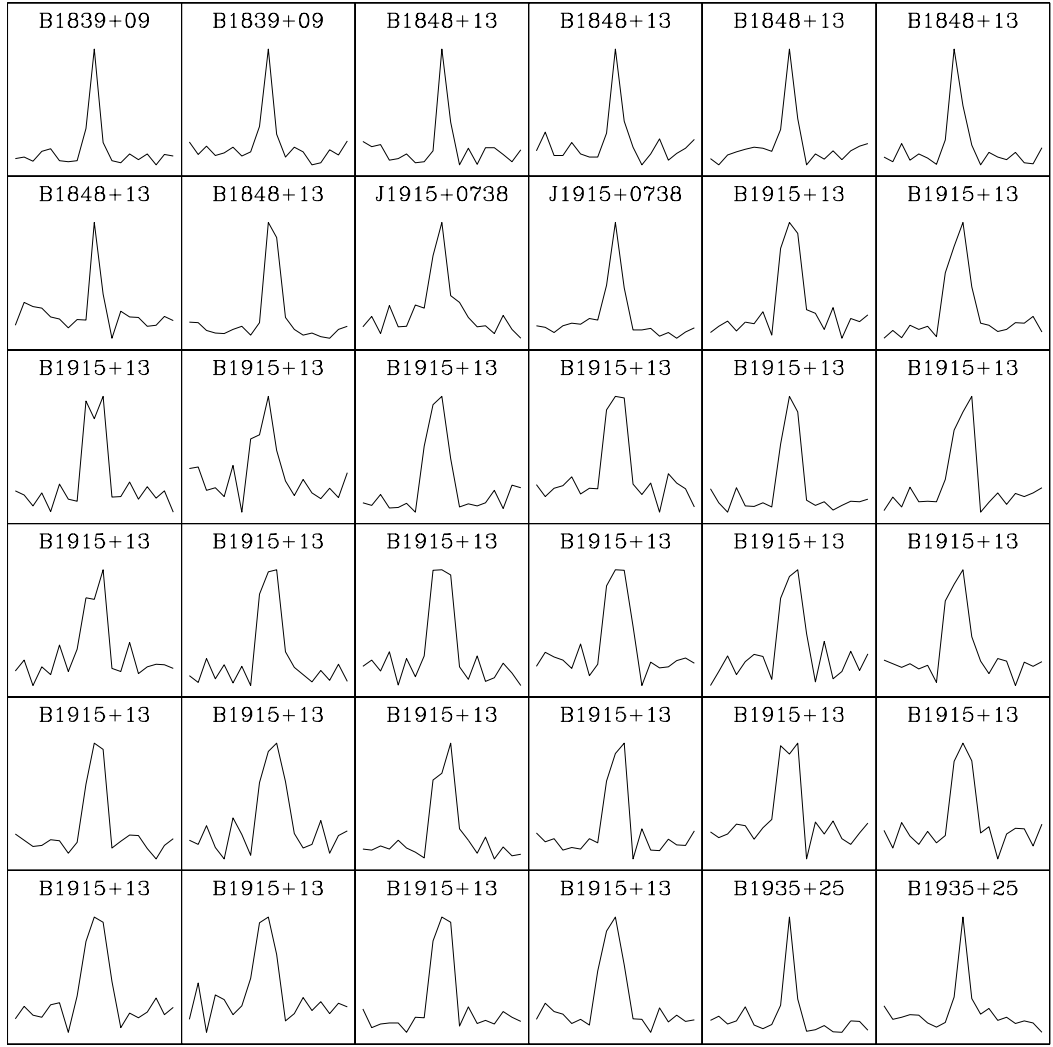}}
\caption{\label{fig:pulses} Thirty-six pulses detected from previously
known pulsars.  Each plot has resolution 8\,ms and covers 160\,ms, centered
on the pulse.}
\end{figure}

Thirty-seven pulses were detected, nearly all at the lowest time resolution
(8\,ms).  Of these pulses, 36 originated from previously known pulsars
(figure~\ref{fig:pulses}).
Table~\ref{tab:psr} lists the five sources from which these pulses originated,
along with all other known pulsars within $5'$ of a pointing center (the
half-power beam radius).  Catalog pulse strengths given in the table are the
product of the reported 400\,MHz flux density and the pulsar period from the
catalog of Taylor {\it et al.} (1995)\nocite{tmlc95}.

The detections and non-detections of known pulsars are reasonably consistent
with the detection threshold of $42\times10^{-4}$\,Jy\,s for 8\,ms pulses,
given that scintillation by the interstellar medium can substantially increase
or decrease pulsar flux densities for a given observation. Sensitivity to some
of these pulsars was degraded by long periods or high duty cycles; for
example, PSR\,J1906+1854 has a pulse width of 64\,ms, a full 8 times longer
than our lowest time resolution.  In principle such long pulses would be
better detected by the further reducing the time resolution of the data, but
in practice distinction of dispersed pulses from non-dispersed radio frequency
interference becomes increasingly difficult.

In addition to signals from known pulsars, one pulse not associated with any
known source was detected.  This pulse was detected with signal-to-noise ratio
25, dispersion measure 24\,pc\,cm$^{-3}$, at time resolution 8\,ms, when
pointing towards $\alpha(2000)=19^{\rm h}18^{\rm m}32^{\rm s}$,
$\delta(2000)=08^\circ38'32''$.  Manual examination of the data found the
pulse width to be 5\,ms and the dispersion measure to be closer to
28\,pc\,cm$^{-3}$.  No other dispersed pulses were detected in this pointing,
and no statistically significant periodicities were detected.  However,
follow-up observations found a pulsar at this position.

\section{PSR J1918+08}\label{sec:1918}

The position of the pulse was observed again on 2 July 1998, using
the Arecibo telescope at 433\,MHz.  The Penn State Pulsar Machine recorded
128-channel spectra across an 8\,MHz passband with 80\,$\mu$s
time resolution over an integration time of 1200\,s.  Inspection of the data
revealed the presence of several strong pulses, and an underlying periodicity
of 2.1\,s, indicative of a radio pulsar.  

\begin{figure}
\center{\ \plotone{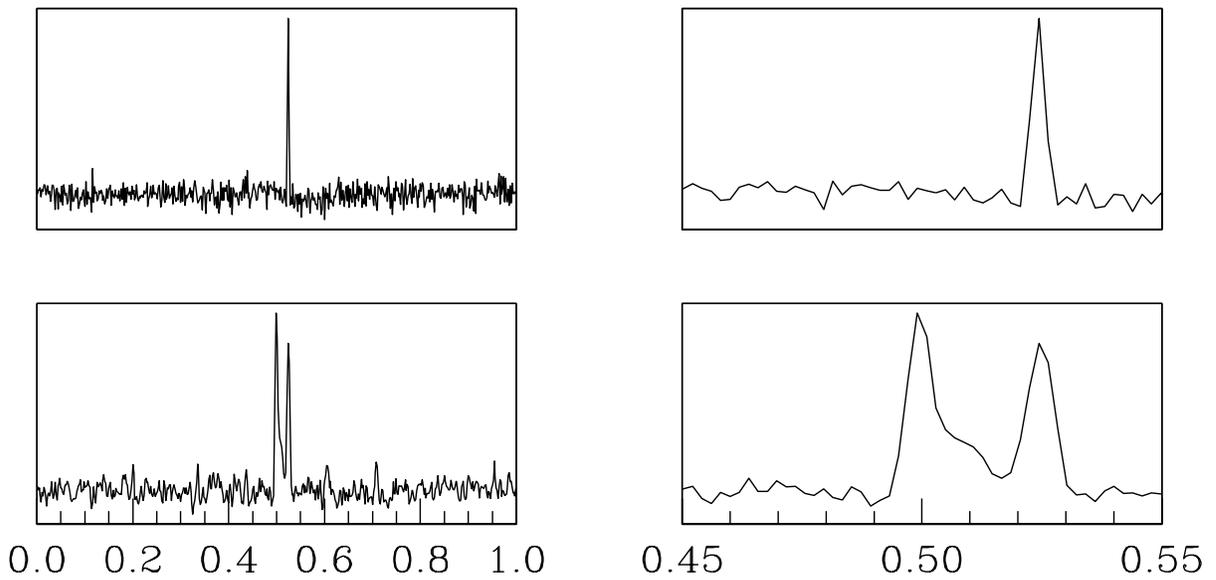}}
\caption{\label{fig:profile} Pulse profile of PSR~J1918+08.
Lower left: Average profile calculated by summing all clean data.
Lower right: Closeup of peak.
Upper left, right: The ``discovery pulse'' showing its alignment
with the average profile.}
\end{figure}

Further observations were made on 21 and 22 July 1998.  Each data set was
folded modulo the pulsar period.  A pulse profile calculated by summing all of
the data is shown in figure~\ref{fig:profile}. The average flux density is
roughly 0.8\,mJy (due to scintillation and crude calibration procedures this
is accurate to no better than a factor of 2).

Standard techniques were used to calculate and analyze pulse times of arrival
from these data.  By splitting the data into two sub-bands and measuring the
arrival time offset, the dispersion measure was determined to be $30\pm
1$\,pc\,cm$^{-3}$.  A phase connected timing solution spanning 2--22 July 1998
yielded a pulse period of $2.1296640\pm0.0000004$, corrected to the solar
system barycenter. The uncertainty in the period is primarily due to
uncertainty in the position, which is only known to the size of the telescope
beam, $\pm 5'$.

With knowledge of the pulsar period from the follow-up data, it was possible
to fold the original search data at the pulsar period.  The pulsar is clearly
visible, with the expected pulse shape.  The discovery pulse is aligned
with the trailing component of the pulse profile (figure~\ref{fig:profile}).

\begin{figure}
\center{\ \plotone{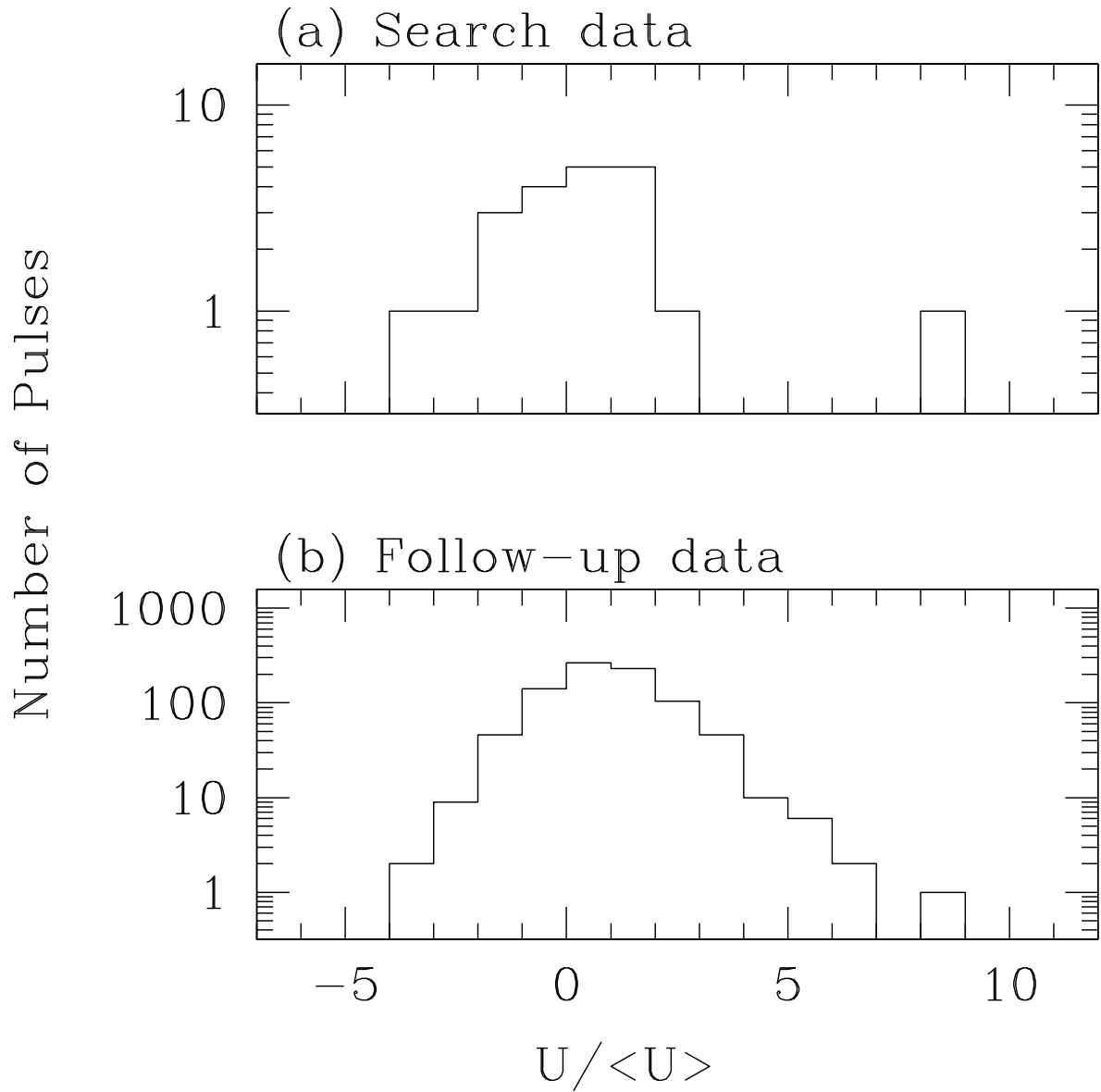}}
\caption{\label{fig:single} Histograms of single pulse strengths.
(a) Original search data.  (b) Follow-up data.}
\end{figure}

To discern whether PSR\,J1918+08 has unusual intensity fluctuations, we
analyzed single pulse strengths from the search and follow-up data.  Strengths
were measured by comparing on-pulse power (phases 0.49 to 0.53 in
figure~\ref{fig:profile}) with off-pulse baseline (phases 0.00 through 0.31
and 0.71 through 1.00.)  Since scintillation is expected to modulate the
observed flux on a time scale of minutes, the strength of each pulse, $U$, was
normalized to the average strength of the 20 nearest pulses in the data set,
$\langle U \rangle$.  Only clean data with sufficiently strong average pulse
signal were included.  The results are summarized in figure~\ref{fig:single}.
The bulk of the pulses are not individually detectable, hence the wide
distribution around zero.  The pulse detected in the original search data had
8 times the average pulse energy.  Based on the much larger sample of pulses
collected in the follow-up data, it appears that such strong pulses are
rare---perhaps one in several hundred pulses---and are the high end of a
smooth distribution of pulse strengths.  Similar intensity distributions, with
occasional pulses as much as 10 times the mean strength, are a common feature
of pulsars (\cite{hw74}).  Thus PSR~J1908+08 is a slow but apparently ordinary
pulsar, and its discovery was made possible by the fortuitous emission of a
rare, strong pulse during the search observation.

\section{Single pulses analysis as a tool for detecting fast pulsars}
\label{sec:analysis}

It is intriguing that, many years after the discovery of the first millisecond
pulsar, no pulsars with period shorter than 1.6\,ms have been detected.
While this may be indicative of a short-period cutoff in the pulsar
population, there are also severe observational biases against such a
detection.  Traditional pulsar searches, using periodic analysis, inevitably
loose sensitivity at short periods due to a combination of instrumental and
interstellar effects.  Here we consider limitations on the detectability of a 
1\,ms periodic pulsar signal in our survey data.

{\it Sample interval.}  The interval at which spectra were sampled was
516\,$\mu$s, so the Nyquist frequency of the data is about 1\,kHz.  Thus, a
1\,ms pulsar would be at the edge of detectability.  In principle, this
restriction could be removed by sampling spectra at shorter intervals.

\begin{figure}
\center{\ \plotone{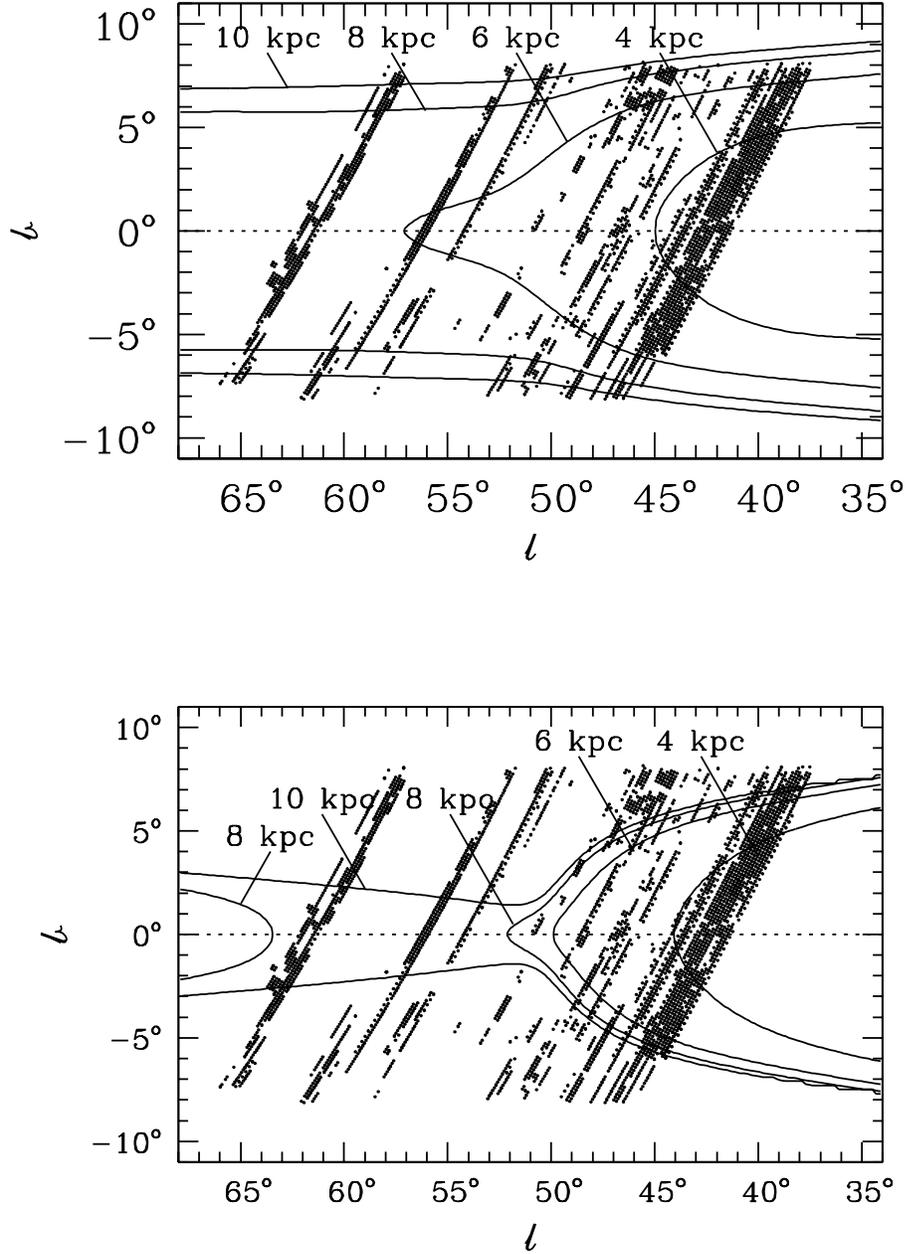}}
\caption{\label{fig:ismlimit} Top---Distance at which differential 
dispersion delay is 1\,ms.  Bottom---Distance at which multipath
scattering broadening is 1\,ms.}
\end{figure}

{\it Dispersion smearing.}  The bandwidth of the individual spectral channels
was 78.125\,kHz.  At the center frequency of 430 MHz, differential dispersion
delays within a single channel broaden the signal by (8.16\,$\mu$s)(DM) where
DM, the dispersion measure, is in pc\,cm$^{-3}$.  Thus for smearing to be less
than 1\,ms, the pulsar must have DM$<123$\,pc\,cm$^{-3}$.
Figure~\ref{fig:ismlimit} shows the distance at which DM=123\,pc\,cm$^{-3}$ in
the region of our search, according to the interstellar electron density model
of Taylor \& Cordes (1993)\nocite{tc93}.  The electron density is highest
towards the inner portion of the Galaxy, restricting the detection of a 1\,ms
pulsar to a distance of no more than 4\,kpc.  The effect of dispersion
smearing could, in principle, be reduced by narrowing the bandwidth of the
individual spectral channels, or using coherent dedispersion techniques, in
either case increasing the dispersion measure at which a 1\,ms pulsar could be
detected.

{\it Scattering.}  Broadening of signals by multipath scattering provides a
more stringent limit on the shortest detectable period.
Figure~\ref{fig:ismlimit} shows the distance at which the scattering time
scale is 1\,ms at 430\,MHz, again using the Taylor \& Cordes
(1993)\nocite{tc93} model.  The distance limit from scattering is also no more
than 4\,kpc at the lowest Galactic longitudes surveyed.  This is a firm
distance limit on {\it any} search for periodic signals towards these
directions at 430\,MHz, and
cannot be surmounted by refinements of the data acquisition system.  Thus very
short period pulsars beyond this distance can only be detected through
unconventional means, such as the single pulse search, or by observing at a
substantially higher radio frequency (at which pulsar flux densities are
typically much lower.)

Of course, the detection of a pulsar through single pulses is possible only if
at least one single pulse is above the detection threshold in a given
observation.  Here we consider the pulse strengths of two well-studied
sources, the 1.6\,ms pulsar B1937+21 and the 33\,ms Crab pulsar, in the
context of our search.

PSR~B1937+21 has a flux density of 230\,mJy at 430 MHz (\cite{ffb91}), and a
period of 1.558\,ms.  Its profile exhibits both a main pulse (MP) and an
interpulse (IP).  Cognard~{\it et al.} (1996)\nocite{cstt96} find that the
relative strengths of the MP and IP are about 5 to 3, and each exhibits giant
pulses with the cumulative probability distribution $f=f_0(U/\langle U
\rangle)^\alpha$, where $f$ is the fraction of pulses or interpulses with
normalized strength greater than $U/\langle U \rangle$, and
$f_0=0.032$ and
$\alpha=-1.8$.  The mean pulse strength of the MP and the IP are
(5/8)(230\,mJy)(1.558\,ms) = 2.2$\times 10^{-4}$\,Jy\,s and
(3/8)(230\,mJy)(1.558\,ms) = 1.3$\times 10^{-4}$\,Jy\,s, respectively.

A single search pointing of 67.715\,s encompasses 43,500 rotations from this
source.  To estimate the strength of the strongest pulse seen in such an
observation, we set $f=1/43500$ and solve for $U/\langle U \rangle=56$.  This
corresponds to a MP strength of $123\times 10^{-4}$\,Jy\,s, well above the
single pulse detection threshold of our search.  Using our highest threshold
for detection of single pulses, $42\times 10^{-4}$\,Jy\,s, and the distance to
PSR\,B1937+21, 3.6\,kpc, we estimate that a similar pulsar would be detectable
to a distance of order $d=(123/42)^{1/2}(3.6)=6.2$\,kpc.  Thus the single
pulse analysis is sensitive to such pulsars at distances somewhat beyond the
scattering limit of the periodic search. 
The distance limit and survey coverage, $\Omega=68\,$deg$^2$, combine to give
the volume searched for 1937+21-like pulsars to be $\Omega d^3/3\sim
1.6$\,kpc$^{3}$.  Since no such pulsars were detected, an upper limit on the
space density of such systems is of order 0.6\,kpc$^{-3}$.

PSR~B0531+21, the Crab pulsar, has a flux density of 650\,mJy at 400\,MHz
(\cite{lylg95}) and a period of 33.4\,ms.  In 800\,MHz observations, Lundgren
{\it et al.} (1995)\nocite{lcu+95} found that 2.5\% of the pulses are
``giant'' pulses, with pulse strength about 20 or more times the mean.
Assuming the pulse strength distribution is similar at 430\,MHz, giant pulses
would have pulse strengths of at least (20)(650\,mJy)(33.4\,ms)=4300$\times
10^{-4}$\,Jy\,s, about 1000 times our detection threshold.  Since the Crab
pulsar undergoes 2000 rotations during a single search pointing, there would
almost certainly be a giant pulse emitted, and it would be well above the
detection threshold for any distance within the Galaxy.

We conclude that the single pulse search is a potentially powerful method of
searching for fast pulsars like B0531+21 and B1937+21 in relatively distant
portions of our Galaxy.  The lack of detections of fast pulsars in our survey
can be attributed to the modest volume covered by our observations, and the
relatively low space density of luminous, short period pulsars.

\acknowledgements

We thank J.~H.~Taylor and A.~S.~Fruchter for collaborating in the data
collection, D.~Copeland for contributions towards the analysis software, and
A.~Wolszczan, S.~B.~Anderson, and B.~J.~Cadwell for making the Penn State
Pulsar Machine available.  The Arecibo Observatory is a facility of the
National Astronomy and Ionosphere Center, operated by Cornell University under
a cooperative agreement with the National Science Foundation.  Pulsar research
at Princeton is supported by NSF grant AST~9618357.

\pagebreak


\pagebreak

\clearpage

\clearpage

\begin{deluxetable}{cccccc}
\tablewidth{0pt}
\tablecaption{\label{tab:param}Search Parameters}
\tablehead{
\colhead{Samples} &
  \colhead{Time} & 
  \colhead{$(S/N)_{\rm min}$}  & 
  \colhead{Flux Density}   &  
  \colhead{Pulse Strength}  &
  \colhead{Dispersion Measure} \\
\colhead{Summed} &
  \colhead{Resolution} & 
  &
  \colhead{Threshold}   &  
  \colhead{Threshold}  &
  \colhead{Range} \\
&  
  \colhead{(ms)}   & 
  & 
  \colhead{(Jy)} & 
  \colhead{($10^{-4}$\,Jy\,s)} & 
  \colhead{(pc\,cm$^{-3}$)} 
}
\startdata
\phn1  &  0.5  &  \phn9  &  1.40  & \phn7 &  \phn2$-$127\phn \nl
\phn2  &  1.1  &  \phn9  &  0.99  &    10 &  \phn2$-$255\phn \nl
\phn4  &  2.1  &  \phn9  &  0.70  &    14 &  \phn6$-$511\phn \nl
\phn8  &  4.1  &     11  &  0.61  &    25 &  \phn8$-$1021 \nl
   16  &  8.3  &     13  &  0.51  &    42 &     16$-$2042 \nl
\enddata
\end{deluxetable}

\clearpage

\begin{deluxetable}{lcccc}
\tablecaption{\label{tab:psr} Previously known pulsars within $5'$ of pointing
positions.}
\tablehead{
\colhead{PSR} & \colhead {Period} & \colhead{Catalog Pulse Strength} &
   \colhead {Number of Pulses} & \colhead {Notes} \\ & \colhead {(s)} &
   \colhead{($10^{-4}$\,Jy\,s)} & \colhead {Detected} & \\ }
\startdata
B1839+09    &  0.381  &    76    & \phn2 &   \nl
J1848+0823  &  0.329  & \phn9    &       &   \nl
B1848+13    &  0.346  &    21    & \phn6 &   \nl
J1902+9723  &  0.488  & \phn3    &       &   \nl
B1906+09    &  0.830  &    42    &       & 1 \nl
J1906+1854  &  1.019  &    47    &       &   \nl
B1911+09    &  1.242  &    37    &       & 1 \nl
B1913+10    &  0.404  &    24    &       &   \nl
J1915+0738  &  1.543  &    29    & \phn2 &   \nl
B1915+13    &  0.195  &    84    &    24 &   \nl
J1916+07    &  0.542  &    22    &       & 1 \nl
J1918+0734  &  0.212  & \phn7    &       &   \nl
B1922+20    &  0.238  &    10    &       & 1 \nl
B1925+18    &  0.483  &    14    &       &   \nl
B1929+20    &  0.286  &    49    &       &   \nl
B1935+25    &  0.381  &    13    & \phn2 &   \nl
J1941+1026  &  0.905  &    16    &       &   \nl
J2008+2513  &  0.589  &    16    &       &   \nl
J2019+2425  &  0.004  & \phn1    &       &   \nl
\enddata
\tablecomments{1.  Pulsar not detected in periodic signal search of
Paper I.}
\end{deluxetable}

\end{document}